\documentclass[aps, prx, twocolumn,show pacs,show keys,superscript address,superscript reference,floatfix]{revtex4-1}
\usepackage[colorlinks=true,citecolor=blue,linkcolor=blue,urlcolor=blue]{hyperref}
\usepackage[sort&compress]{natbib}
\usepackage{graphicx,times}
\usepackage{color}
\usepackage{amssymb}

\usepackage{subeqnarray}
\usepackage{float}
\usepackage{bbm}
\usepackage{bm}
\usepackage{diagbox}
\usepackage{makecell}
\usepackage{amsmath}

\DeclareMathAlphabet\mathbfcal{OMS}{cmsy}{b}{n}
\newcommand{\ket}[1]{\ensuremath{\left|{#1}\right\rangle}}
\newcommand{\bra}[1]{\ensuremath{\left\langle{#1}\right|}}

\begin{document}
\title{Tripartite entanglement in quantum memristors}
\author{S. Kumar}
\affiliation{International Center of Quantum Artificial Intelligence for Science and Technology (QuArtist) \\
and Physics Department, Shanghai University, 200444 Shanghai, China} 

\author{F.~A.~C\'ardenas-L\'opez}
\affiliation{International Center of Quantum Artificial Intelligence for Science and Technology (QuArtist) \\
and Physics Department, Shanghai University, 200444 Shanghai, China} 

\author{N.~N.~Hegade}
\affiliation{International Center of Quantum Artificial Intelligence for Science and Technology (QuArtist) \\
and Physics Department, Shanghai University, 200444 Shanghai, China} 

\author{F.~Albarr\'an-Arriagada}
\email{ pancho.albarran@gmail.com}
\affiliation{International Center of Quantum Artificial Intelligence for Science and Technology (QuArtist) \\
and Physics Department, Shanghai University, 200444 Shanghai, China} 

\author{E.~Solano}
\email{enr.solano@gmail.com}
\affiliation{International Center of Quantum Artificial Intelligence for Science and Technology (QuArtist) \\
and Physics Department, Shanghai University, 200444 Shanghai, China} 
\affiliation{IKERBASQUE, Basque Foundation for Science, Plaza Euskadi, 5, 48009 Bilbao, Spain}
\affiliation{Kipu Quantum, Kurwenalstrasse 1, 80804 Munich, Germany}

\author{G.~Alvarado~Barrios}
\email{phys.gabriel@gmail.com}
\affiliation{International Center of Quantum Artificial Intelligence for Science and Technology (QuArtist) \\
and Physics Department, Shanghai University, 200444 Shanghai, China} 

\begin{abstract}

We study the entanglement and memristive properties of three coupled quantum memristors. We consider quantum memristors based on superconducting asymmetric SQUID architectures which are coupled via inductors. The three quantum memristors are arranged in two different geometries: linear and triangular coupling configurations. We obtain a variety of correlation measures, including bipartite entanglement and tripartite negativity. We find that, for identical quantum memristors, entanglement and memristivity follow the same behavior for the triangular case and the opposite one in the linear case. Finally, we study the multipartite correlations with the tripartite negativity and entanglement monogamy relations, showing that our system has genuine tripartite entanglement. Our results show that quantum correlations in multipartite memristive systems have a non-trivial role and can be used to design quantum memristor arrays for quantum neural networks and neuromorphic quantum computing architectures.
\end{abstract}

\maketitle

\section{Introduction}

\label{sec.1}

Neuromorphic computing is the brain-inspired computational paradigm where digital or analog systems mimic neural systems~\cite{Schuman2017}. Its most prominent structures are artificial neural networks \cite{Nielsen2015}, which have shown remarkable breakthrough applications in recent years \cite{Ren2017}. A fundamental element in analog artificial neural networks is the memristor. This device was proposed by L. Chua in 1971 \cite{IEEE.1971} as the fourth circuit element and was only experimentally realized in 2008 by HP labs~\cite{Nat.2008}. The relevance of memristors for analog artificial neural networks lies in their memory effects, and nonlinear dynamics, which are similar to neural synapses \cite{Wang2018NM}. Furthermore, memristor-enabled neuromorphic computing goes beyond the von Neumann computing paradigm, avoiding the von Neumann bottleneck, one of the most fundamental limits of current computers~\cite{Li2018JPD,Markovic2020NatRevPhys,Islam2019JPD}. 
  
On the other hand, quantum computing \cite{Gyongyosi2019CompSciRev} aims to revolutionize computation by harnessing uniquely quantum phenomena to surpass the capabilities of classical computers, with remarkable recent breakthroughs~\cite{Arute2019Nature,Zhong2020Science,Wu2021PRL}, including digital-analog quantum neural networks~\cite{GongarXiv2022}. The fusion of quantum computing and neuromorphic computing is known as neuromorphic quantum computation (NQC)~\cite{Markovica2020}, which aims to implement brain-inspired devices with quantum hardware and software, and may lead to new groundbreaking technologies. It is natural to wonder if a quantum version of memristors, i.e. quantum memristors (QMs), could become a fundamental component of neuromorphic quantum hardware similar to its classical counterpart. This question has motivated the theoretical proposal of quantum memristors in different platforms, such as in quantum photonics ~\cite{sanz2018,materials.2020} and superconducting circuits~\cite{SciRep.2016, PRapp.2016, PRapp.2018,Phys.Rev.Applied.2014,Sci.Rep.2016}, and remarkably, has seen recent experimental realizations ~\cite{Spagnolo.2021,Gao2020arXiv}.

Naturally, to utilize quantum memristors for NQC applications, it is necessary to understand how they can be connected and correlated. However, most proposals have been limited to single quantum memristor models, and further study is required for the case of coupling and scaling quantum memristors. In this context, Kumar et al.~\cite{Shubham.PRA} have studied memristive dynamics of two coupled quantum memristors concerning their entanglement properties. In that work, they find that the memristive properties and entanglement have an opposite behavior, providing the first signs of a non-trivial role of quantum correlations in memristive devices. Nonetheless, there is more to explore in this direction since increasing the number of subsystems introduces new quantum correlations that cannot be observed in bipartite systems, such as multipartite entanglement and entanglement monogamy~\cite{Coffman2000,Wang2014PRL}. Addressing this point can be relevant for future applications, which will require scaling quantum memristor arrays into quantum neural networks and neuromorphic quantum computers. 

\begin{figure}[b]
\centering
\includegraphics[width=0.6\linewidth]{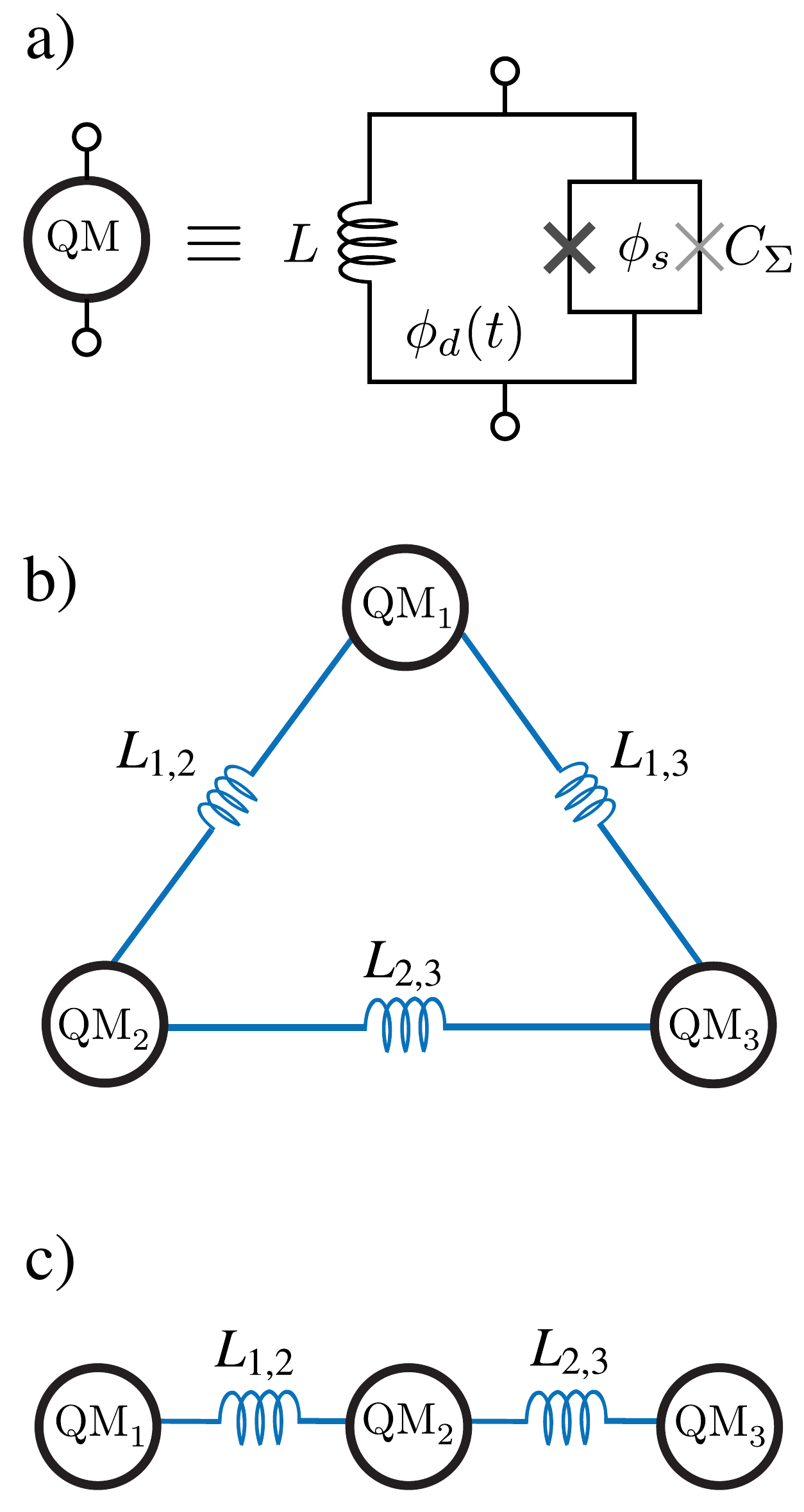}
\caption{(a) Superconducting QM composed of a CA-SQUID with effective capacitance, $C_{\Sigma}$, coupled to an inductor, $L$, in a loop threaded by a time-dependent, flux $\phi_{d}(t),$, and a static signal, $\phi_{s}$, through the SQUID. (b) Triangular and (c) linear coupling via inductors. }
\label{Fig01}
\end{figure}

In this work, we study the case of three coupled QMs in two different geometries, a linear and a triangular configuration. We focus on the relation between the non-trivial entanglement properties of tripartite systems and the memristive dynamics of the system. Since our system can be partitioned in several different ways, we consider entanglement of formation (EoF)~\cite{PhysRevLett.80.2245} to quantify bipartite quantum correlations. In addition, we quantify the genuine tripartite entanglement using the tripartite negativity \cite{Euro.Phys.2008}, and entanglement monogamy relations. We find that the hysteretic behavior of the QMs can follow the same or opposite behavior as the quantum correlations, depending on the coupling geometry. This work helps to understand the role of multipartite quantum correlations in coupled quantum memristors in developing NQC and bio-inspired technology.

This article is organized as follows, In Sec.~\ref{Sec02}, we describe the three superconducting QMs model used in our work. In Sec.~\ref{Sec03}, we describe the performance of the three QMs proposal in terms of the form factor, the entanglement of formation of the different bipartitions, tripartite negativity, and monogamy relation to evaluating the global correlations in the system. In Sec.~\ref{Sec04}, we provide the main conclusions of our work. Finally, in the Appendices, we provide the technical details on the derivation of the Hamiltonian (Appendix \ref{AppA}) and the memristive equations (Appendix \ref{AppB}).
\\
\section{The model}
\label{Sec02}

We consider the quantum memristor model from the theoretical proposal based on a superconducting conductance-asymmetric SQUID (CA-SQUID), presented in Refs.~\cite{Phys.Rev.Applied.2014,Sci.Rep.2016} and shown here in Fig.~\ref{Fig01} a. Therefore, a quantum memristor is composed of a CA-SQUID connected in parallel to an inductance, where the loop formed between the SQUID and the inductor is threaded by a magnetic flux, $\phi_d(t)$. The CA-SQUID refers to a SQUID composed of different Josephson junctions (JJs), which have different conductance. In this device, by fixing the flux inside the SQUID ($\phi_s$) to half a flux quantum, the critical current of each JJ can be cancelled. Then, we have a dominant current contribution corresponding to the quasiparticle current, which is dissipative and memristive by nature~\cite{Martinis2009PRL}. Since the critical current cancels, each QM can be described by a simple harmonic oscillator Hamiltonian
\begin{equation}
\hat{\mathcal{H}}_j = E_{C_j} \hat{n}_j^{2} + \frac{E_{L_j}}{2} \hat{\phi}_j^{2}=\hbar \omega_j a_j^{\dagger}a_j,
\label{Eq01}
\end{equation}
where the subindex $j$ refers to the $j$th QM. Here, $\hat{n}_j$ and $\hat{\phi}_j$ are the dimensionless number and phase operators, respectively. $E_{C_j}=2e^2\hat{C}_{\Sigma_j}^{-1}$ and $E_{L_j}=\varphi_{0}^{2}\hat{L}_j^{-1}$ are the charging and inductive energy of the QM, and $\varphi_{0} = \hbar/2e$ is the reduced flux quantum. Also $\omega_j = \sqrt{2E_{C_j}E_{L_j}}/\hbar$ is the harmonic oscillator frequency, $\hat{n}_j = \frac{i}{2\eta_j} (a_j^{\dagger} - a_j)$ and $\hat{\phi}_j= \eta_j (a_j^{\dagger} + a_j)$. Here, 
\begin{equation}
\eta_j=\sqrt[4]{\frac{E_{C_j}}{2E_{L_j}}}
\end{equation}
and $a_j^{\dagger}(a_j)$ is the creation(anihilation) operator.  For simplicity, we consider from now $\hbar=1$.

As the system undergoes quasiparticle-induced decay it is described by the following master equation
\begin{eqnarray} 
\label{Master_equation_1_memristor}
\frac{d}{dt}\hat{\rho}_j(t) = &&-i \big[ \hat{\mathcal{H}}_j,{\hat{\rho}_j} \big]\nonumber\\
&& + \sum_{\ell=1}^{3}\frac{\Gamma_j(t)}{2} \bigg[ a_j \hat{\rho}_j a_j^{\dag} - \frac{1}{2} \{ a_j^{\dag} a_j, \hat{\rho}_j\} \bigg],
\end{eqnarray}
where $\Gamma_j(t)$ is the time-dependent decay rate, given by ${\Gamma_j(t) = \lvert \bra{0}\sin(\hat{\phi}_j/2)\ket{1}\lvert^{2}S_{\rm{qp}}(\omega)}$, where $S_{\rm{qp}}(\omega)$ is the spectral density of the quasiparticle bath~\cite{Catelani2011prl}. Using the  flux quantization condition on the outer loop of the memristor, we can write the decay rate in terms of the external magnetic flux $\phi_d(t)$ as $\Gamma_j(t) = g_j^{2} \omega \exp(-g_j^{2}) ({1 + \cos[\phi_d(t)]})/2$. The memristive behavior appears in the dynamics of the quasiparticle current and the voltage in the capacitor in the CA-SQUID $I_{\textrm{qp}}=\Gamma(t)V_{\textrm{cap}}$, which are obtained from the equations of motion for $\hat{n}_j$ and $\hat{\phi}_j$,
\begin{eqnarray} 
\frac{d}{dt} \langle \hat{n}_j \rangle &=& E_{L_j} \langle \hat{\phi}_j \rangle - \frac{\Gamma_j(t)}{2} \langle \hat{n}_j \rangle, \nonumber \\
\frac{d}{dt} \langle \hat{\phi}_j \rangle &=& -E_{C_j} \langle \hat{n}_j \rangle - \frac{\Gamma_j(t)}{2}  \langle \hat{\phi}_j \rangle ,
\end{eqnarray}
obtaining a phase-dependent memristive relations.  

For the case of three inductively coupled quantum memristors, the Hamiltonian of the system is given by (see Appendix~\ref{AppA})
\begin{eqnarray}
\label{H_2_mem}
\hat{H} = &&\sum_{\ell=1}^3\hat{\mathcal{H}}_j- \sum_{j<k=1}^3 E_{L_{j,k}}\hat{\phi}_{j}\hat{\phi} _{k} , \nonumber \\
=&&\sum_{\ell=1}^3\hat{\mathcal{H}}_j- \sum_{j<k=1}^3 g_{j,k}({a_j-a_j^\dagger})({a_k-a_k^\dagger}) .
\end{eqnarray}

\begin{figure*}[t]
\centering
\includegraphics[width=0.95\linewidth]{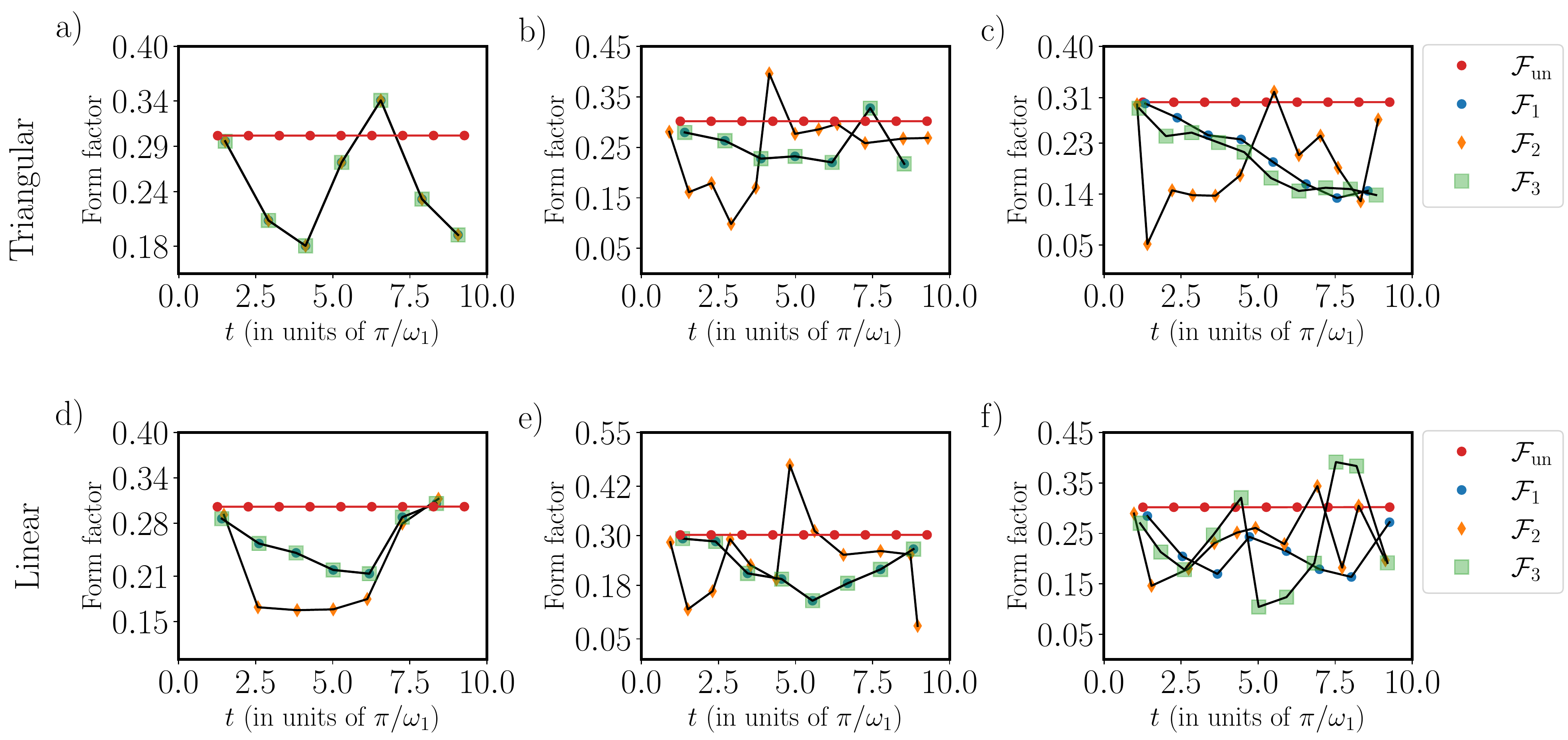}
\caption{Form factor dynamics. Upper panels correspond to triangular configuration and lower panels to the linear case. We consider three cases for each configuration,  these are: identical memristors  $QM_{1} = QM_{2} = QM_{3}$ for a) triangular configuration, with parameters, $C_{\Sigma_{1}}=C_{\Sigma_{2}}=C_{\Sigma_{3}}= $ 3.6$~[\rm{fF}]$, $L_{1}=L_{2}=L_{3}= $ 0.2$~[\rm{\mu H}]$ $L_{1,2}=L_{2,3}=L_{1,3}=2~[\rm{\mu H}]$, and d) linear configuration, with parameters, $C_{\Sigma_{1}}=C_{\Sigma_{2}}=C_{\Sigma_{3}}= $ 3.6$~[\rm{fF}]$, $L_{1}=L_{2}=L_{3}= $ 0.2$~[\rm{\mu H}]$ $L_{1,2}=L_{2,3} =2~[\rm{\mu H}]$ and $L_{1,3}=0$. For one non-identical memristor $QM_1=QM_3\ne QM_2$ for b) triangular configuration with parameters, $C_{\Sigma_{1}}=C_{\Sigma_{3}}= $ 3.6$~[\rm{fF}]$, $C_{\Sigma_{2}}= $ 2.6$~[\rm{fF}]$, $L_{1}=L_{2}=L_{3} =0.2~[\rm{\mu H}]$, $L_{1,2}=L_{2,3}=1.69~[\rm{\mu H}]$ and $L_{1,3}= 2~[\rm{\mu H}]$, and e) linear configuration with parameters, $C_{\Sigma_{1}}=C_{\Sigma_{3}}= $ 3.6$~[\rm{fF}]$, $C_{\Sigma_{2}}= $ 2.6$~[\rm{fF}]$, $L_{1}=L_{2}=L_{3} =0.2~[\rm{\mu H}]$, $L_{1,2}=L_{2,3}=1.69~[\rm{\mu H}]$ and $L_{1,3}= 0$. Lastly, all non-identical memristors, $QM_1\ne QM_2\ne QM_3$, for c) triangular configuration with parameters, $C_{\Sigma_{1}}= $ 3.6$~[\rm{fF}]$, $C_{\Sigma_{2}}= $ 2.6$~[\rm{fF}]$, $C_{\Sigma_{3}}= $ 3$~[\rm{fF}]$, $L_{1}=L_{2}=L_{3}= $ 0.2$~[\rm{\mu H}]$, $L_{1,2}=1.69~[\rm{\mu H}]$, $L_{2,3}=1.55~[\rm{\mu H}]$ and $L_{1,3}=1.83~[\rm{\mu H}]$, and for f) linear configuration with paremeters, $C_{\Sigma_{1}}= $ 3.6$~[\rm{fF}]$, $C_{\Sigma_{2}}= $ 2.6$~[\rm{fF}]$, $C_{\Sigma_{3}}= $ 3$~[\rm{fF}]$, $L_{1}=L_{2}=L_{3}= $ 0.2$~[\rm{\mu H}]$, $L_{1,2}=1.69~[\rm{\mu H}]$, $L_{2,3}=1.55~[\rm{\mu H}]$ and $L_{1,3}=0$. The initial state for each QM for all cases was $\ket{\Psi(\pi/4,\pi/2)}$.}
\label{Fig02}
\end{figure*}


Here, $ E_{L_{j,k}} =\varphi_{0}^{2}\hat{L}_{j,k}^{-1}$ is the inductive energy of the inductor among the $j$th and $k$th QM. The coupling strength  $g_{j,k}$ reads
\begin{eqnarray}
g_{j,k}=\frac{\sqrt{L_jL_k}}{L_{j,k}} \sqrt{\omega_j\omega_k}\Rightarrow E_{L_{j,k}}=\frac{\sqrt[4]{2}}{2}\frac{g_{j,k}}{\eta_j\eta_k}.
\end{eqnarray}

We consider two different geometries for our system, the triangular one, where all the coupling strengths are different from zero (see Fig.~\ref{Fig01} b), and the linear one, which means that $g_{1,3}=0$ (see Fig.~\ref{Fig01} c). Now, the time-dependent master equation for the three QMs system reads

\begin{equation} 
\frac{d}{dt}\hat{\rho}(t)= -i \big[ \hat{H},{\hat{\rho}} \big] + \sum_{j=1}^{3}\frac{\Gamma_j(t)}{2}\left[a_j \hat{\rho} a_j^{\dag} - \frac{1}{2} \{ a_j^{\dag}a_j,\hat{\rho}\}\right].
\label{Eq07}
\end{equation}

Using Eq.~\ref{Eq07}, we can obtain the equations of motion for $\langle \hat{n}_j \rangle$ and $\langle \hat{\phi}_j \rangle$, which are given by (see Appendix~\ref{AppB})

\begin{subequations}
\begin{eqnarray} \nonumber
\label{current1}
\frac{d}{dt}\langle\hat{n}_j \rangle = && E_{L_j} \langle \hat{\phi}_j \rangle - \frac{{\Gamma }_j(t)}{2}\langle \hat{n}_j \rangle\nonumber\\
&& -(E_{L_{j,k}}\langle \hat{\phi}_{k} \rangle+ E_{L_{j,l}} \langle \hat{\phi}_l \rangle ), \\
\frac{d}{dt}\langle{\hat{\phi}}_{j}\rangle = && -2E_{C_{j}}\langle{\hat{n}}_j\rangle - \frac{\Gamma _{j}(t)}{2}\langle{\hat{\phi}}_{j}\rangle ,
\end{eqnarray}
where $j\ne k\ne l$. As $E_{L_{j,k}}\sim g_{j,k}$, for the linear coupling, we have that $g_{1,3}=E_{L_{1,3}}=0$.
\end{subequations}

To study the correlations and performance induced by the coupling, we start with a maximal voltage state for all the QMs, meaning initial states for the $j$th QM  of the form  $\ket{\Psi_{j}(\theta_{j},\varphi_{j})} = \cos(\theta_{j}/2)\ket{0}+e^{i\varphi_{j}}\sin(\theta_{j}/2)\ket{1}$, with $\varphi_j = \pi/2$. And starting from the initially separable state for the three QMs $\ket{\Psi_1} \otimes \ket{\Psi_2}\otimes\ket{\Psi_3}$.

\section{Performance} 
\label{Sec03}
\subsection{Form factor}
The principal characteristics of a QM are its quantumness and its hysteresis loop. It is known that the area enclosed by this curve can be related to the memory effects of a QM. Also, the physical properties and the maximum values of the voltage (input variable) are related to the perimeter of the hysteresis curve. In this line, a good candidate to evaluate the memristive properties of the memristor $j$ is the form factor~\cite{Shubham.PRA} given by
\begin{equation}
\mathcal{F}_j=4\pi\frac{A_j}{P_j^2}, 
\end{equation}
where $A_j$ and $P_j$ are the area and perimeter of the pinched hysteresis loop of the memristor $j$. We note that the form factor is a dimensionless quantity which measures the area enclosed by a given perimeter, where its maximal value $\mathcal{F}=1$ is obtained for a circle (maximal area). Meanwhile, the minima value $\mathcal{F}=0$ for a line (minimal area). It allows us to compare the different curves without caring about the losses in the maximum voltage value in each loop.

\begin{figure}[t]
\centering
\includegraphics[width=0.85\linewidth]{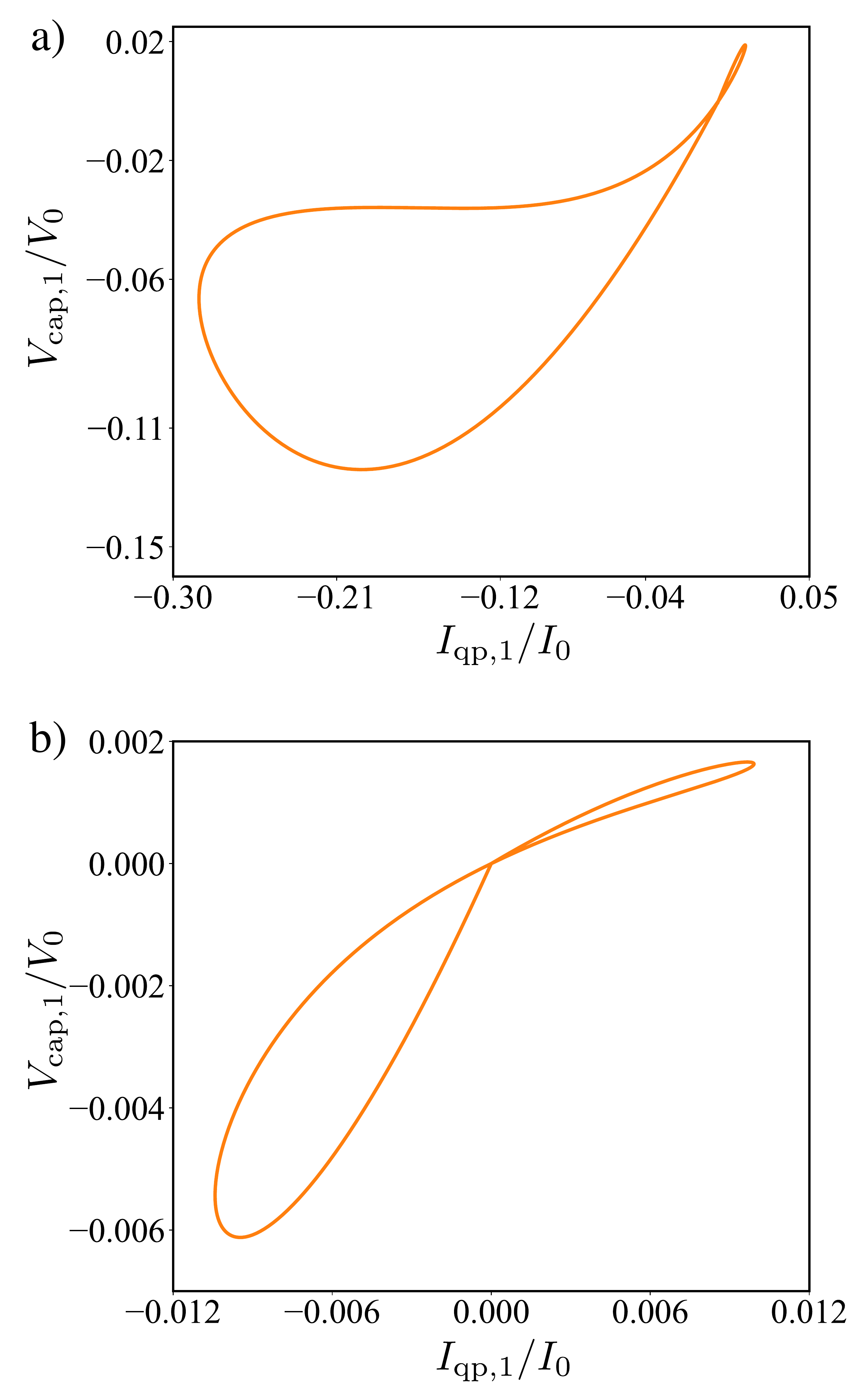}
\caption{Pinched hysteresis curve for the largest a) and smallest b) form factor reported in Fig.~\ref{Fig02} e).}
\label{Fig03}
\end{figure}

We consider the three possible cases: first, when all three QMs are identical ($\textrm{QM}_1=\textrm{QM}_2=\textrm{QM}_3$), second, two of them identical ($\textrm{QM}_1=\textrm{QM}_3\ne\textrm{QM}_2$), and third when all of them are nonidentical ($\textrm{QM}_1\ne\textrm{QM}_2\ne\textrm{QM}_3$). We define the initial and final point of the hysteresis curve when it crosses the origin; then, a hole loop is defined the voltage-current curve crosses the origin two times. Figure~\ref{Fig02} shows the form factors of the three cases mentioned above for the triangular and linear coupling. For the uncoupled case, all the memristors have the same form factor, which is a constant in time (horizontal red line), which means that the shape of the hysteresis loop is the same for all the uncoupled memristor at any time but with different scales. Also, for all the cases, this figure shows that the coupling introduces oscillations in the form factor, with regions that surpass the uncoupled case. A notable result is that Fig. \ref{Fig02} e has a form factor larger to $0.5$ corresponding to the maximal value for the form factor for symmetric pinched loops corresponding to two joined circles. It means that high values in the form factor, induced by the coupling, implies large asymmetry between the lobes of the memristive curve as we can see in Fig.~\ref{Fig03} a, which shows the memristive curve for the highest form factor in Fig.~\ref{Fig02} e. It allows surpassing the maximal memory properties in symmetric hysteresis curves. Also, from Fig.~\ref{Fig03} e, we obtain the minimal value for the form factor, which implies that the lobes tend to shrink. This case of minimal form factor is shown in Fig.~\ref{Fig03} b, where the lobes decrease their area, implying a reduction of the memristive properties. Linear coupling allows obtaining maximal memory properties during the evolution, improving the previous enhancement in two coupled memristors reported in Ref.~\cite{Shubham.PRA}.

\subsection{Quantum correlations}
\begin{figure*}[t]
\centering
\includegraphics[width=0.95\linewidth]{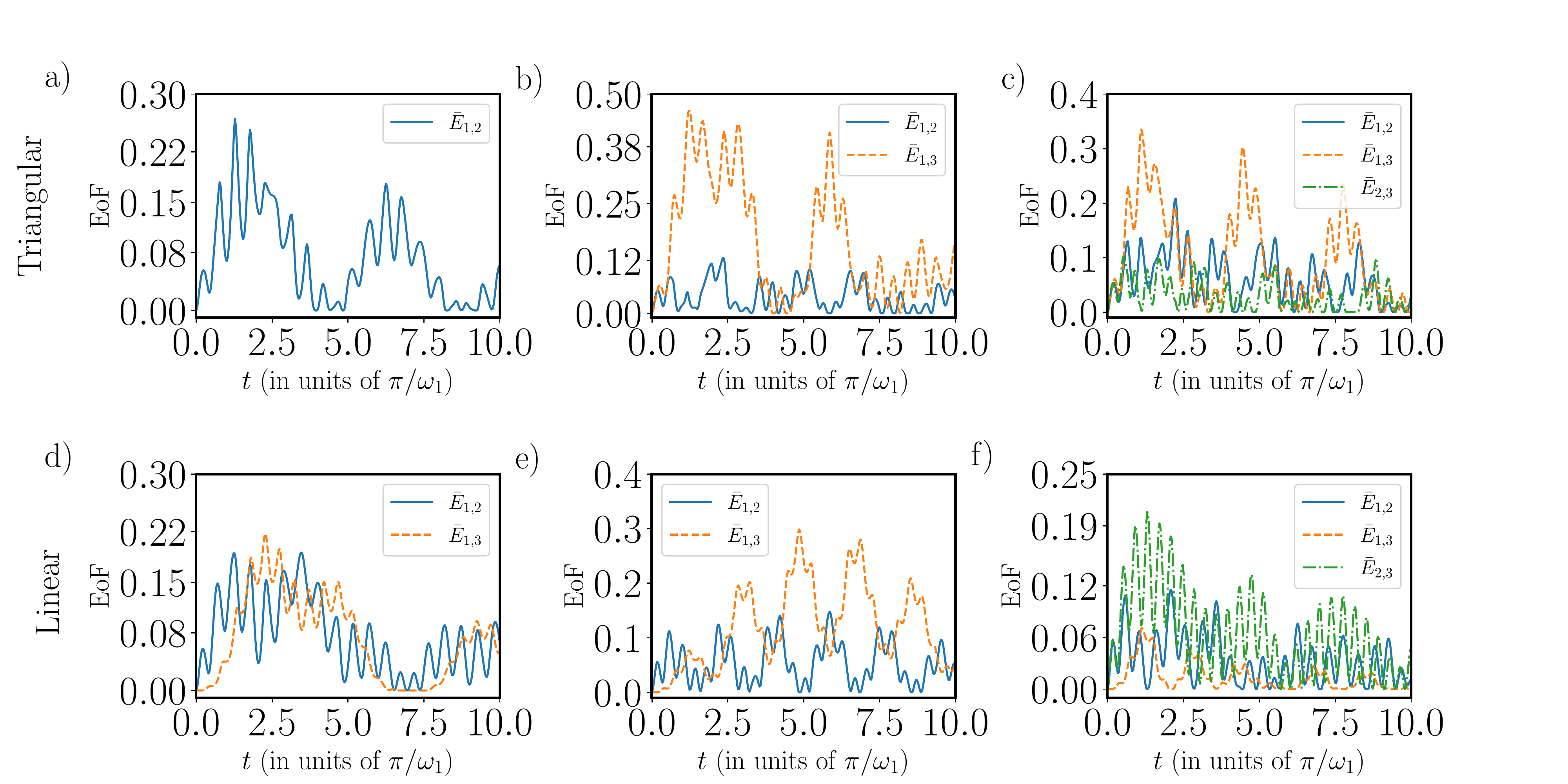}
\caption{Bipartite concurrence and form factor for triangular (upper panels), linear (lower panels) coupled QMs. We have used the same initial state and system parameters as in Fig.~\ref{Fig02}.}
\label{Fig04}
\end{figure*}
We characterize the quantum correlations in our system using entanglement of formation (EoF) in the different non-equivalent pairs of QMs. Also, we consider the multipartite correlations given for the tripartite negativity for the entire system. We finish our study by considering the entanglement monogamy relation to show that our coupled QMs have correlations beyond bipartitions.

\textit{Entanglement of Formation.--}  First, we consider the correlations in the reduced system of two QMs given by the trace of one QM in the total system. In this case, the density matrix is given by
\begin{equation}
\rho_{j,k}(t)=\textrm{Tr}_{l}[\rho(t)] ,
\end{equation}
where $j\ne k\ne l$ are the label of the three QMs. 

Now, we can calculate the EoF of a mixed state as
\begin{equation}
\bar{E}(\rho)=\min_\mathcal{K}\sum_jp_j^{(\mathcal{K})}E(|\phi_j^{(\mathcal{K})}\rangle),
\label{EoF}
\end{equation}
where $p_j^{(\mathcal{K})}$ and $|\phi_j^{(\mathcal{K})}\rangle$ are the $j$th probability and the $j$th state of the pure state decomposition $\mathcal{K}$ of the density matrix $\rho$. It means
\begin{equation}
\rho=\sum_j p_j^{(\mathcal{K})}|\phi_j^{(\mathcal{K})}\rangle\langle\phi_j^{(\mathcal{K})}|,
\end{equation}
as the states $|\phi_j^{(\mathcal{K})}\rangle$ and $|\phi_k^{(\mathcal{K})}\rangle$ can be non-orthogonal, then the decomposition is not unique and have the extra index $\mathcal{K}$. Here, the pure state bipartite entanglement $E$ reads
\begin{equation}
E(\ket{\phi})=-\textrm{Tr}[\rho_{1(2)}\log_d\rho_{1(2)}] ,
\end{equation}
where $\rho_{1(2)}=\textrm{Tr}_{2(1)}[\ket{\phi}\bra{\phi}]$ is the partial trace over one of the subsystems of the pure state, and $d$ the dimension of $\rho_{1(2)}$. In this sense, the EoF given by Eq.~(\ref{EoF}) is the minimization of the average entanglement over all the possible pure state decomposition of the mixed state $\rho$. 

As each QM is described effectively by a $2\times2$ matrix (qubit), the EoF of the reduce two QMs systems have an analytical formula which reads
\begin{equation}
\bar{E}(\rho)=-h(\rho)\log_2[h(\rho)] - [1-h(\rho)]\log_2[1-h(\rho)] ,
\end{equation}
with
\begin{equation}
h(\rho)=\frac{1+\sqrt{1-C(\rho)}}{2} .
\end{equation}
Here, $C(\rho)$ is a quantity called concurrence that for a two-qubit matrix $\rho$ is given by
\begin{equation}
C=\textrm{max}\{0,2\lambda_{max}-\textrm{Tr}(R)\} ,
\end{equation}
where $R=\sqrt{\rho^{1/2}\tilde{\rho}\rho^{1/2}}$ with maximal eigenvalue given by $\lambda_{max}$, and $\tilde{\rho}=(\sigma_y\otimes\sigma_y)\rho^*(\sigma_y\otimes\sigma_y)$, with $\rho^*$ the conjugate matrix of $\rho$.

Figure~\ref{Fig04} shows the EoF for the different non-equivalent pairs of QMs for linear and triangular configuration. In this figure, $\bar{E}_{j,k}$ is the EoF between the QMs $j$ and $k$. We can also see that for identical QMs case, the form factor and the EoF follow the same oscillatory behavior in the triangular configuration, which means that they have maximum (minimum) values in the same time region. On the other hand, for linear configuration for identical QMs, the form factor and EoF follow opposite dynamics, where the maximum of one curve coincide with minimal of the other curve, obtaining a similar conclusion that for the two coupled QMs in Ref.~\cite{Shubham.PRA}. Nevertheless, this similar behavior is lost in the other cases.

Also, we can observe that $\bar{E}_{1,3}$ presents regions with zero entanglement for linear configuration for Fig.~\ref{Fig04} d and f. This phenomenon is commonly known as sudden death and sudden birth of entanglement, a characteristic of memory dynamics. We observe that the dynamics of the EoF present two-time scales ($\tau_{1(2)}$), given by the competition between the input signal for the $j$th QM given by $\tau_1=2\pi/\omega_j$, and the coupling effect between the QMs $j$ and $k$ which reads $\tau_2=2\pi/g_{j,k}$.

\begin{figure*}[!t]
\centering
\includegraphics[width=0.95\linewidth]{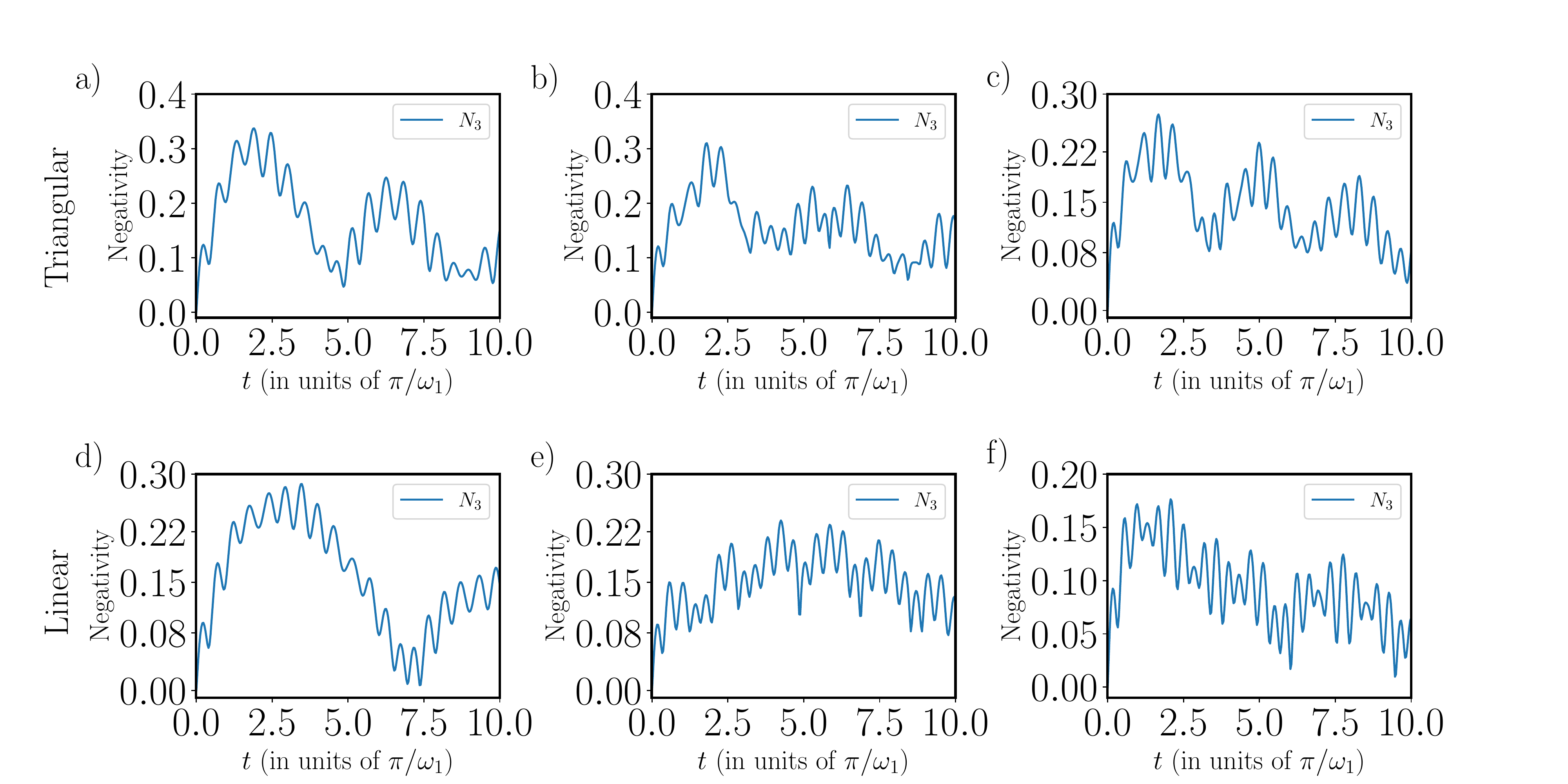}
\caption{Negativity for triangular and linear geometry for $QM_1=QM_2=QM_3$ (a and d), $QM_1=QM_3\ne QM_2$ (b and e) and $QM_1\ne QM_2\ne QM_3$ (c and f). We have used the same initial state and system parameters as in Fig.~\ref{Fig02}.}
\label{Fig05}
\end{figure*}

\begin{figure*}[!t]
\centering
\includegraphics[width=1\linewidth]{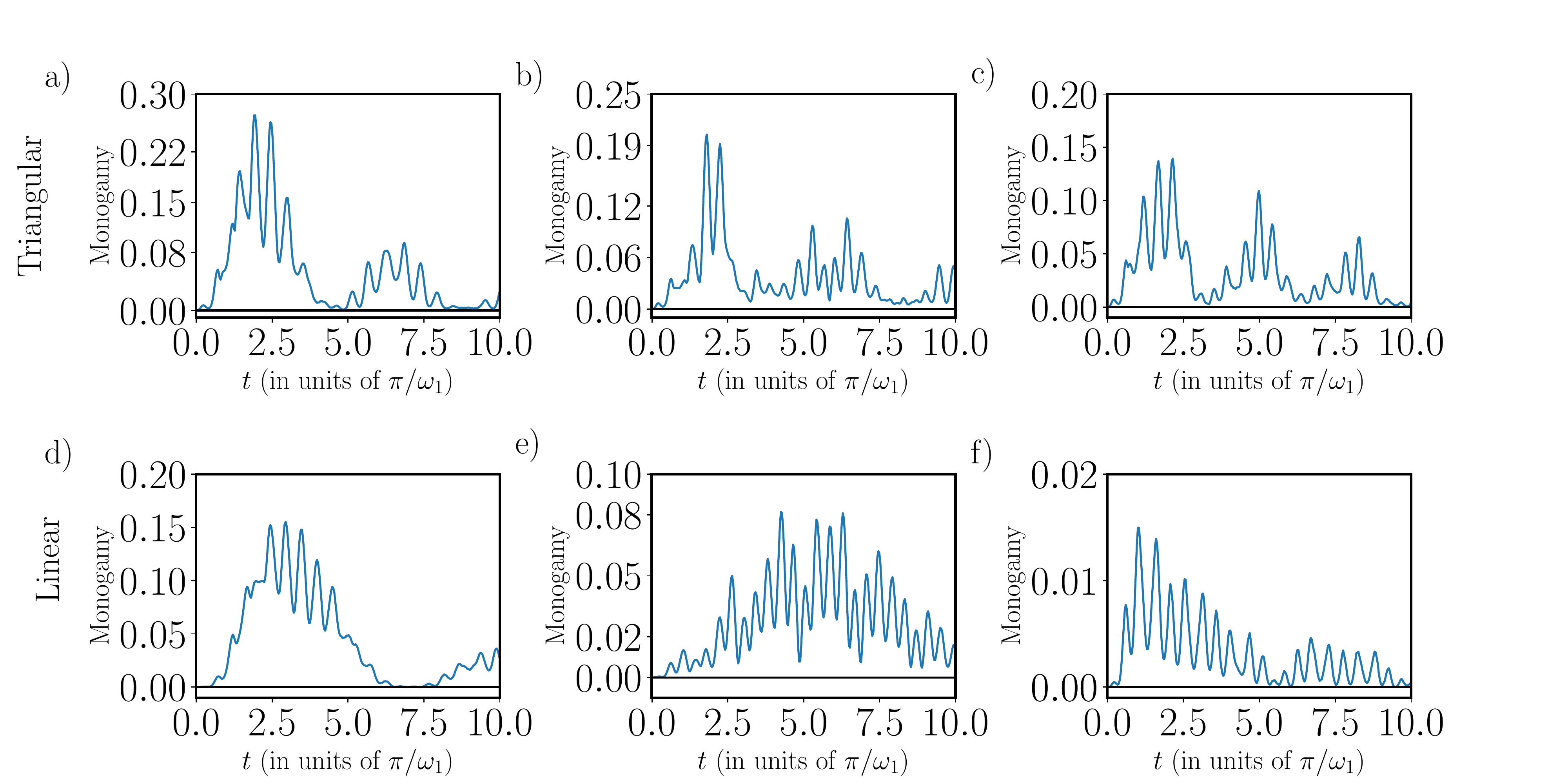}
\caption{The interplay between monogamy of entanglement of formation and the form factor for (a) triangular and (b) linear coupling. The system parameters and initial states are chosen to be the same as for the cases of triangular and linear coupling in Fig.~\ref{Fig02}. }
\label{Fig06}
\end{figure*}

\textit{\textit{Tripartite correlations.--}} Now, to measure the degree of genuine tripartite correlations, we use the tripartite negativity defined as
 \begin{equation}
 \mathcal{N}_3(\rho) = (\mathcal{N}_{1-23}\mathcal{N}_{2-13}\mathcal{N}_{3-12})^{1/3} ,
 \end{equation}
 where
\begin{equation}
\mathcal{N}_{j-kl}=-2\sum_n\sigma_n(\rho^{T_j})
\end{equation}
 is the bipartite negativity. Here, $\sigma_n(\rho^{T_j})$ is the $n$th negative eigenvalue of the the matrix $\rho^{T_j}$ which represents the partial transpose of the density matrix $\rho$ respect to the subsystem $j$. $\rho^{T_j}$ and $\rho$ are related by
 \begin{equation}
\bra{\alpha_j,\beta_{kl}} \rho^{T_j}\ket{\gamma_j,\delta_{kl}}=\bra{\gamma_j,\beta_{kl}} \rho^{T_j}\ket{\alpha_j,\delta_{kl}},
 \end{equation} 
where the subindex indicates each of the two parts of the bipartition. This quantity is an entanglement witness based on the Peres-Horodecki criterion~\cite{Peres1996PRL,Horodecki1996PLA}, where a $\mathcal{N}_3\ne0$ implies that the quantum state cannot be written as a bipartition of its constituent subsystems. Figure~\ref{Fig05} shows the negativity in time for the different cases under study. We can observe that the tripartite negativity for $t>0$ is always different from zero, which means that the three QMs in all the cases under study have multipartite entanglement. This can also be seen from the monogamy entanglement relation with respect to QM$_2$, which reads
\begin{equation}
\bar{E}_{2,13}^2\geq\bar{E}_{2,1}^2+\bar{E}_{2,3}^2 .
\end{equation}
This relation is saturated when the EoF between $QM_2$ and the joint system form by the $QM_1$ and $QM_3$ correspond only to bipartite entanglement. Therefore, the next multipartite entanglement measure can be defined as
\begin{equation}
\mathcal{M}_2=\bar{E}_{2,13}^2-\bar{E}_{2,1}^2-\bar{E}_{2,3}^2 ,
\end{equation}
which is different from zero if we have entanglement beyond the entanglement of its pairs. Figure~\ref{Fig06} shows the value of $\mathcal{M}_2$ in time for all the cases under study. We can observe that we have $\mathcal{M}_2=0$ only for the case of identical QMs in the linear geometry. This result is expected since there is no direct coupling between $QM_1$ and $QM_3$. Nevertheless, it is interesting that for non-identical QMs, this behavior disappears, obtaining EoF beyond bipartitions.

As we observe form the numerical results, the three coupled QM system shows entanglement properties beyond the case of two coupled QM, maintaining similar conclusions only for the three identical QM case. Beyond identical case, we can observe interesting behaviors like high form factor and genuine tripartite quantum correlations. Also we need to mentions that in this system the switching of the hysteresis loop can be observed, but its relation with the entanglement properties is not clear as in the two coupled QM case reported in Ref. \cite{Shubham.PRA}.

\section{Conclusions}
\label{Sec04}
We have studied the entanglement and memristive properties of a three QMs system in different configurations. Specifically, we analyze the EoF for different bipartitions and tripartite negativity for the whole system. Also, we analyze the entanglement monogamy relation to show when the entanglement properties go beyond bipartite EoF. We find that the EoF for different bipartitions can be easily related with the form factor dynamics only for identical QMs. In such cases, we obtain the same behavior for triangular coupling, i.e., that the maximal values of EoF and form factor correspond to the same time region, and opposite behavior for the linear case. This implies that the maximal value of one quantity corresponds to the same time that the other get the minimal one. According to the tripartite negativity, the system is in a genuine tripartite entangled quantum state at any time. The monogamy relation also shows that only a few cases do not exist EoF beyond the entanglement among individuals QMs. On the other hand, linear coupling allows reaching high values of form factor, which means strong memory properties, surpassing the values in Ref.~\cite{Shubham.PRA} for two entangled QMs.

This work opens the door to studying the role of quantum correlations and memristive features in more complex networks of QMs suitable for the analog implementation of neuromorphic quantum computing. Also, showing the possibility to engineer the memory properties of QMs, a crucial feature for the development of quantum neural networks.

\section{Acknowledgements}
The authors acknowledge support from NSFC~(12075145) and Shanghai~STCSM~(Grant 2019SHZDZX01-ZX04).

\onecolumngrid
\appendix
\appendix

\section{Derivation of the circuit Hamiltonian}
\label{AppA}
The circuit Lagrangian for the most general tripartite quantum memristor reads
\begin{eqnarray}\nonumber
\mathcal{L} = && \sum_{\ell=1}^3\bigg[\frac{C_{\Sigma,\ell}}{2}\dot{\varphi}_{\ell}^{2} + E_{J\ell}\cos\bigg(\frac{\varphi_{\ell}+\phi_{d\ell}(t)}{\varphi_{0}}\bigg) + E_{J\ell}\cos\bigg(\frac{\varphi_{\ell}+\phi_{d\ell}(t)+\phi_{s\ell}}{\varphi_{0}}\bigg)-\frac{\varphi_{\ell}^{2}}{2L_{\ell}}\bigg]\\
&&  - \frac{(\varphi_{2}-\varphi_{1})^{2}}{2L_{c1}}- \frac{(\varphi_{3}-\varphi_{2})^{2}}{2L_{c2}}- \frac{(\varphi_{1}-\varphi_{3})^{2}}{2L_{c3}},
\end{eqnarray}
where, $C_{\Sigma,\ell}$ stands to the effective capacitance of the $\ell$th CA-SQUIDs, $E_{J\ell}$ is the Josephson energy of the $\ell$th junction, $\phi_{s\ell}$ is the static magnetic flux threading the inner loop forming the CA-SQUID (see Fig.~\ref{Fig01}{(a)}). Moreover, $L_{\ell}$ is the inductance of the $\ell$th QM, and  $\phi_{d\ell}(t)$ is the time-dependent external magnetic flux threading the outer loop of each QM (see Fig.~\ref{Fig01}{(a)}). Finally, $\varphi_0=\hbar/2e$ is the reduced quantum magnetic flux. When the external flux in the inner loop satisfies $\phi_{s\ell}/\varphi_{0}=\pi$, we neglect each energy contribution of the Josephson junctions forming the CA-SQUID, leading to the simplified Lagrangian
\begin{eqnarray}
\mathcal{L} = \sum_{\ell=1}^3\bigg[\frac{C_{\Sigma,\ell}}{2}\dot{\varphi}_{\ell}^2 - \frac{\hat{L}^{-1}_{\ell,\ell}}{2}\varphi_{\ell}^{2}\bigg] + \hat{L}^{-1}_{1,2}\varphi_{1}\varphi_{2}+\hat{L}^{-1}_{2,3}\varphi_{2}\varphi_{3} + \hat{L}^{-1}_{1,3}\varphi_{1}\varphi_{3}. 
\end{eqnarray}
Here, $\hat{L}^{-1}_{j,k}$ is the matrix element $(j,k)$ of the inverse of the inductance matrix defined as 
\begin{eqnarray}
\hat{L}^{-1}=\left(
\begin{array}{ccc}
 \frac{1}{L_1}+\frac{1}{L_{c_{1}}}+\frac{1}{L_{c_{3}}} & -\frac{1}{L_{c_{1}}} & -\frac{1}{L_{c_{3}}}\\
 -\frac{1}{L_{c_{1}}}& \frac{1}{L_2}+\frac{1}{L_{c_{1}}}+\frac{1}{L_{c_{2}}} &-\frac{1}{L_{c_{2}}} \\
 -\frac{1}{L_{c_{3}}} &-\frac{1}{L_{c_{2}}}&\frac{1}{L_3}+\frac{1}{L_{c_{2}}}+\frac{1}{L_{c_{3}}} \\
\end{array}
\right).
\end{eqnarray}
We obtain the circuit Hamiltonian by applying the Legendre transformation $\mathcal{H}=\sum_{\ell}q_{\ell}\dot{\varphi}_{\ell}-\mathcal{L}$, where $q_{\ell}=\partial \mathcal{L}/\partial\dot{\varphi_{\ell}}=C_{\Sigma,\ell}\dot{\varphi}_{\ell}$ is the canonical conjugate momenta. Thus, we obtain
\begin{equation}
\mathcal{H} = \sum_{\ell=1}^3\bigg[C_{\Sigma,\ell}^{-1}\frac{q_{\ell}^2}{2}  + \hat{L}^{-1}_{\ell,\ell}\frac{{\varphi_{\ell}^{2}}}{2}\bigg] -\hat{L}^{-1}_{1,2}\varphi_{1}\varphi_{2}-\hat{L}^{-1}_{2,3}\varphi_{2}\varphi_{3} - \hat{L}^{-1}_{1,3}\varphi_{1}\varphi_{3}, 
\end{equation} 
we proceed to quantize the Hamiltonian by promoting the classical variables to quantum operators $q_{\ell}\rightarrow\hat{q}_{\ell}=-2e\hat{n}_{\ell}$, and $\varphi_\ell/\varphi_0\rightarrow\hat{\phi}_{\ell}$, satisfying cannonica commutation relations $[\hat{\phi}_{\ell},\hat{n}_{\ell'}]=i\delta_{\ell,\ell'}$, leading to the quantum Hamiltonian
\begin{eqnarray}
\label{Hamiltonian}
\hat{H}&=&\sum_{\ell=1,2}\bigg[E_{C_{{\ell}}}\hat{n}_{\ell}^{2} +\frac{E_{L_{\ell}}}{2}\hat{\phi}_{\ell}^{2}\bigg] - E_{L_{1,2}}\hat{\phi}_{1}\hat{\phi_{2}} - E_{L_{2,3}}\hat{\phi}_{2}\hat{\phi_{3}} - E_{L_{1,3}}\hat{\phi}_{1}\hat{\phi_{3}},
\end{eqnarray}
where $E_{C_{{\ell}}}=2e^2/C_{\Sigma_{\ell}}$ is the charge energy, $E_{L_{i,j}}=\varphi_{0}^{2}\hat{L}^{-1}_{i,j}$ is the coupling inductive energy, respectively. For simplicity, we can redefine $\hat{n}_{\ell}$ and $\hat{\phi}_{\ell}$ in terms of creation and annihilation operators as
\begin{eqnarray}
\hat{n}_{\ell}&=&\frac{i}{4g_{\ell}}(a_{\ell}^{\dag}-a_{\ell}),\\
\hat{\phi}_{\ell}&=&2g_{\ell}(a_{\ell}^{\dag}+a_{\ell}),
\end{eqnarray}
Here $g_{\ell}=(E_{C_{{\ell}}}/32E_{L_{\ell,\ell}})^{1/4}$ is proportional to the zero-point fluctuation of the phase operator $\hat{\phi}_{\ell}$. Thus, the circuit Hamiltonian reads 
\begin{equation}
\hat{{H}}=\sum_{\ell=1}^3\hbar \omega_\ell\hat{a}_\ell^{\dagger}\hat{a}_\ell - {g_{12}}(\hat{a}_1^\dagger+\hat{a}_1)(\hat{a}_2^\dagger+\hat{a}_2)-{g_{23}}(\hat{a}_2^\dagger+\hat{a}_2)(\hat{a}_3^\dagger+\hat{a}_3)-{g_{13}}(\hat{a}_1^\dagger+\hat{a}_1)(\hat{a}_3^\dagger+\hat{a}_3),
\label{Hamiltonian}
\end{equation}
where $\omega_{\ell}=\sqrt{2E_{C_{{\ell}}}E_{L_{\ell}}}/\hbar$ is the frequency of the $\ell$th QM, $g_{ij}=k_{ij}\sqrt{\omega_{i}\omega_j}$ is the coupling strength between them with $k_{ij}=\frac{\sqrt{L_iL_j}}{L_{i,j}}$ as the ratio between the inductances. Notice that this circuit regards the triangular configuration described in the main manuscript. However, by taking $g_{13}\rightarrow 0$ we obtain the linear coupling case.

\section{Derivation of dynamic coupling equations}
\label{AppB}
In this section, we derive the equation of motion for the observables $\hat{n}_{\ell}$, and $\hat{\phi}_{\ell}$, which are related to the current flowing and the voltage across the memristor. We will consider the dynamics in the Schr\"odinger picture, where it is governed by the following master equation
\begin{eqnarray}
\label{Mst_eq_1}
\frac{d}{dt}\hat{\rho}(t) &=& -i \big[ \hat{H},{\hat{\rho}} \big] + \sum_{j=1}^{3}\frac{\Gamma_j(t)}{2}\left[a_j \hat{\rho} a_j^{\dag} - \frac{1}{2} \{ a_j^{\dag}a_j,\hat{\rho}\}\right].
\end{eqnarray}
Here, $\hat{H}$ is the system Hamiltonian in Eq. (\ref{Hamiltonian}), while $\Gamma_{\ell}(t)= \lvert \bra{0}\sin(\phi_{\ell}/2)\ket{1}\lvert^{2}S_{\rm{qp}}(\omega_{\ell})$ is the decay rate of the quasiparticle tunneling of the $j$th QMs. For the expectation values we obtain that the equation of motions can be written as 
 \begin{eqnarray}
 \label{Chap3_Sec3_Eq3}
\frac{d\langle\hat{\mathcal{O}}(t)\rangle}{dt}= -\frac{i}{\hbar}{\rm{Tr}}\bigg[[\mathcal{H},\mathcal{O}]\hat{\rho}(t)\bigg] +{\rm{Tr}}\bigg[ \tilde{\mathcal{D}}[\mathcal{O}]\hat{\rho}(t)\bigg],
 \end{eqnarray}
where $\tilde{\mathcal{D}}[\mathcal{O}]=\Gamma(t)(a^{\dag}\mathcal{O} a - \{a^{\dag}a,\mathcal{O}\}/2)$ corresponds to the Lindbladian for the operator $\mathcal{O}$. For $\hat{n}_{j}$, we obtain 
\begin{subequations}
\begin{eqnarray}
\label{Chap3_Sec3_Eq4b}\nonumber
\frac{d}{dt}\langle\hat{n}_j \rangle = &&  E_{L_{j}} \langle \hat{\phi}_j \rangle - E_{L_{1,2}}(\delta_{j,1}\langle \hat{\phi}_{2}\rangle+\delta_{j,2}\langle \hat{\phi}_{1}\rangle)- E_{L_{2,3}}(\delta_{j,2}\langle \hat{\phi}_{3}\rangle+\delta_{j,3}\langle \hat{\phi}_{2}\rangle)- E_{L_{1,3}}(\delta_{j,1}\langle \hat{\phi}_{3}\rangle+\delta_{j,3}\langle \hat{\phi}_{1}\rangle)\\
&-& {{\rm{Tr}}}\bigg[\tilde{\mathcal{D}}[\hat{n}_{j}]\hat{\rho}(t)\bigg]
\end{eqnarray}
\end{subequations}
and
\begin{eqnarray}
\tilde{\mathcal{D}}[\hat{n}_{j}]=\frac{4i\Gamma(t)}{g_{0}}\bigg[-\frac{a_{j}^{\dag}}{2} +\frac{a_{j}}{2}\bigg]=-\frac{\Gamma(t)}{2}\hat{n}_{j} \, .
\end{eqnarray}
Thus, the equation of motion reads
\begin{eqnarray}\nonumber
\label{current1}
\frac{d}{dt}\langle\hat{n}_j \rangle = &&  E_{L_{j}} \langle \hat{\phi}_j \rangle - E_{L_{1,2}}(\delta_{j,1}\langle \hat{\phi}_{2}\rangle+\delta_{j,2}\langle \hat{\phi}_{1}\rangle)- E_{L_{2,3}}(\delta_{j,2}\langle \hat{\phi}_{3}\rangle+\delta_{j,3}\langle \hat{\phi}_{2}\rangle)- E_{L_{1,3}}(\delta_{j,1}\langle \hat{\phi}_{3}\rangle+\delta_{j,3}\langle \hat{\phi}_{1}\rangle)- \frac{{\Gamma }_\ell(t)}{2}\langle \hat{n}_j \rangle.\\
\label{voltage1}
\end{eqnarray}
Notice that we can write the equation of motion in a compact form as follows
\begin{eqnarray}\nonumber
\label{current1}
\frac{d}{dt}\langle\hat{n}_j \rangle = && 3E_{L_{j}} \langle \hat{\phi}_j \rangle- \sum_{n,m=1,2,3}\bigg[ \frac{E_{L_{n,m}}}{2}(\delta_{j,n}\langle\hat{\phi}_{m}\rangle+\delta_{j,m}\langle\hat{\phi}_{n}\rangle)\bigg]- \frac{{\Gamma }_\ell(t)}{2}\langle \hat{n}_j \rangle.\\
\label{voltage1}
\end{eqnarray}
Similarly, we obtain the equation of motion for the phase operator $\phi_j$ as follows 
\begin{eqnarray}
\frac{d}{dt}\langle{\hat{\phi}}_{j}\rangle = && -2E_{C_{j}}\langle{\hat{n}}_j\rangle - \frac{\Gamma _{j}(t)}{2}\langle{\hat{\phi}}_{j}\rangle.
\end{eqnarray}

\appendix
\appendix

\end{document}